\documentclass[
  twocolumn,
  usenames,dvipsnames]        
  {aastex631}

  \usepackage{amssymb,amsmath,soul}

  \usepackage[ignoremode, finalnew]{trackchanges} 

  \newcommand{\vct}[1] {\ensuremath{\boldsymbol{#1}}}    
  \newcommand{\comment}[1]{\textcolor{red}{\bf #1}}

  \newcommand{\vB} {\vct{B}}
  \newcommand{\vb} {\vct{b}}
  \newcommand{\vj} {\vct{j}}
  \newcommand{\vv} {\vct{v}}   
  \newcommand{\vx} {\vct{x}}
  \newcommand{\vz} {\vct{z}}
  
  \newcommand{\lc} {\lambda_C}
  \newcommand{\ts} {\lambda_T}
  \newcommand{\di} {d_\mathrm{i}}
  \newcommand{\tb}  {t_b}            
  \newcommand{\Ceps} {C_{\epsilon}}

  \newcommand{\avg}[1] {\langle #1 \rangle}
 
 \renewcommand{\d} {\mathrm{d}}
   \newcommand{\e} {\mathrm{e}}           

\begin{document}

\title{Statistics of Turbulence in the Solar Wind. I.\\What is the Reynolds Number of the Solar Wind?}

\correspondingauthor{Daniel Wrench}
\email{daniel.wrench@vuw.ac.nz}

\author[0000-0002-7463-3818]{Daniel Wrench}
\affiliation{Victoria University of Wellington \\ 
Kelburn, Wellington 6012, New Zealand}

\author[0000-0003-0602-8381]{Tulasi N. Parashar}
\affiliation{Victoria University of Wellington \\ 
Kelburn, Wellington 6012, New Zealand}

\author[0000-0002-2814-7288]{Sean Oughton}
\affiliation{University of Waikato, Hamilton 3240, New Zealand}

\author{Kevin de Lange}
\affiliation{Victoria University of Wellington \\ 
Kelburn, Wellington 6012, New Zealand}

\author[0000-0001-5840-8760]{Marcus Frean}
\affiliation{Victoria University of Wellington \\ 
Kelburn, Wellington 6012, New Zealand}

\begin{abstract}
The Reynolds number, $Re$, is an important quantity for describing a turbulent flow. It tells us about the bandwidth over which energy can cascade from large scales to smaller ones, prior to the onset of dissipation. However, calculating it for nearly collisionless plasmas like the solar wind is challenging. Previous studies have used ``effective'' Reynolds number formulations, expressing $Re$ as a function of the correlation scale and either the Taylor scale or a proxy for the dissipation scale. We find that the Taylor scale definition of the Reynolds number has a sizeable prefactor of approximately \change{50}{27}, which has not been employed in previous works. Drawing from 18 years of data from the Wind spacecraft at 1\,au, we calculate the magnetic Taylor scale directly and use both the ion inertial length and the magnetic spectrum break scale as approximations for the dissipation scale, yielding three distinct $Re$ estimates for each 12-hour interval. Average values of $Re$ range between \change{173,000}{116,000} and \change{6,330,000}{3,406,000}, within the general distribution of past work. We also find considerable disagreement between the methods, with linear associations of between \change{0.43}{0.38} and \change{0.75}{0.72}. Although the Taylor scale method is 
  arguably
more physically motivated, due to its dependence on the energy cascade rate, more theoretical work is needed in order to identify the most appropriate way of calculating effective Reynolds numbers for kinetic plasmas. As a summary of our observational analysis, we make available a data product of 28 years of 1\,au solar wind and magnetospheric plasma measurements from Wind.



\end{abstract}

    \section{Introduction} \label{sec:intro}

Most naturally-occurring plasmas are either observed to be, or believed to be, in a turbulent state. There is significant variation in the parameters of these systems, including the length and time scales, the plasma $\beta$, the turbulent Mach numbers, and the relative size of the system compared to kinetic scales. Many of these systems are in what is called a ``kinetic'' state, where the dynamical length and time scales of interest are comparable to or smaller than the collisional time scales of interest. Astrophysical examples include the solar wind \citep[e.g.,][]{Bruno2013}, accretion disks \citep[e.g.,][]{balbus1998instability}, and the intracluster medium \citep[e.g.,][]{mohapatra2020turbulence}. For these systems, the collisional closures associated with fluid models are no longer applicable (or at least not obviously so). This means one has to resort to higher-order closures for the fluid models, or, in most cases, to a kinetic description of the plasma \citep[e.g.,][]{marsch2006kinetic}.

Turbulence theories utilize dimensionless parameters to categorize various flow regimes. For homogeneous incompressible 
Navier--Stokes turbulence, the most important of these is the Reynolds number $Re$, defined as the ratio of the characteristic magnitudes of the non-linear inertial term and the viscous term of the Navier--Stokes momentum equation \citep[e.g.,][]{pope_2000}. Herein we define it by
\begin{equation}\label{eq:re_hydro}
    Re = \frac{U\lc}{\nu},
\end{equation}
where $U = \sqrt{ \avg{  \vv \cdot\vv } } $ is the characteristic root-mean-square (rms) speed of the fluctuations,
$\lc$ is the correlation scale (aka outer scale), and $\nu$ is the (kinematic) viscosity; $\vv(\vx, t)$ is the velocity field. Loosely, $\lc$ corresponds to the largest separation at which turbulent fluctuations remain correlated, which in a hydrodynamic context can be thought of as the size of the energy-containing eddies. (It is also often written as $L$ and called the ``characteristic length'' scale.) Small $Re$ implies that the viscous effects are significant and hence the nonlinear term is weak and  will not introduce significant nonlinearities into the system's evolution.  Conversely, a large value of $Re$ implies that the nonlinear term plays a significant role in the dynamics of the fluid. 

This dynamic can be appreciated more clearly when $Re$ is expressed solely in terms of length scales.
One way this can be done is to introduce the Kolmogorov dissipation scale (aka inner scale) 
  $ \eta = (\nu^3/\epsilon)^{1/4} $,
where 
    $ \epsilon  = \nu 
       \avg{  \left( \nabla \times \vv \right)^2
       } 
    $ 
is the mean rate of kinetic energy dissipation 
    \citep{Kol41a,tennekes1972}.
A physical interpretation is that the Kolmogorov scale is where the smallest eddies in the fluid become critically damped, due to their nonlinear (aka turnover) timescale being equal to their dissipation timescale.
Recall also that the dissipation rate   
 can be phenomenologically modelled as 
 \begin{align}
    \epsilon_\text{phenom}
     = 
      \Ceps \frac{U^3}{ \lc} ,
   \label{eq:eps-phenom}   
 \end{align} 
 where $ \Ceps $ is treated as a fitting constant
  \citep[e.g.,][]{Batchelor1970,tennekes1972}.
 Employing this in the definition of $\eta$ 
 yields the form

\begin{equation}\label{eq:re_kolmogorov}
    Re  
    \equiv Re_\eta 
    =  \frac{1}{\Ceps^{1/3}}
          \left( \frac{\lc}{\eta} \right)^{\frac{4}{3}} ,
\end{equation} 
revealing that $Re$ is a measure of the bandwidth of the turbulent energy cascade.
A large $Re$ indicates there is a large separation between the outer and inner scales. 
This larger bandwidth implies there are more scales where the nonlinear term is strong enough to create turbulent structures and thereby increase the intermittency of the flow 
  \citep[see, e.g.,][]{MattEA15-philtran,Parashar2019, Cuesta2022_intermittent}. 
A small   bandwidth, and hence a small   $Re$, implies that dissipation occurs very quickly and damps any turbulent structures that the nonlinear term might try to create. 
Such low-$Re$ 
situations are sometimes seen in planetary magnetosheaths \citep{CzykowskaAnGeo2001,HadidApJL15,HuangApJL17,Chhiber2018}.

Estimating $Re$ for hydrodynamical systems, using Eq.~\eqref{eq:re_hydro}, is straightforward as all the required quantities are well defined and often readily determined in experiments. For kinetic plasmas such as the solar wind, however, it is not possible to write a Chapman-Enskog-like closure to define a viscosity \citep{ChapmanCowlingBook, HuangBook}. (Some attempts have been made to estimate the viscosity of kinetic systems; see e.g., \citet{verma1996, zhuravleva2019, BandyopadhyayEA23-clike, yang2023}. 
This lack of a well-defined viscosity also precludes using Eq.~\eqref{eq:re_kolmogorov}, as it means we cannot define $\eta$. Typically, in kinetic plasmas, one must therefore resort to defining an \emph{effective} Reynolds number. Some hydrodynamic studies have investigated estimating the energy input into the system \citep{Zhou_2014}, as well as using more precise boundaries of the inertial range \citep{Zhou2007, ZhouThornber16}, in order to get around this lack of a clearly-defined inner scale. Herein we describe two approaches to formulating an effective Reynolds number.

\begin{figure} 
    \centering
    \includegraphics[width=\columnwidth]{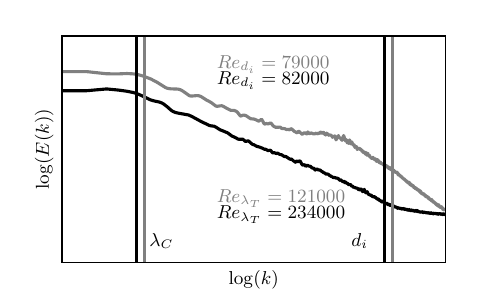}
    \caption{Power spectra (energy $E(k)$ vs.\ wavenumber $k$) for two spacecraft data intervals with \change[]{identical}{very similar} outer and inner scales but different power levels. Vertical lines indicate the respective correlation scales $\lc$ and ion inertial lengths $\di$. On the basis of the power levels one would expect different turbulent behaviour from these intervals. However, using $\di$\remove[]{(or $f_b$)} as the inner scale in Eq.~\eqref{eq:re_kolmogorov} \change[]{would imply they have the same $Re$.}{implies they have almost the same $Re$. Using Eq.~\eqref{eq:re_taylor} we do capture this difference, giving very different values of $Re_{\ts}$}.}
    \label{fig:shifted_spectra}
\end{figure}

The first approach is to apply Eq.~\eqref{eq:re_kolmogorov} and use a different small scale --- one that is observationally calculable --- as a signifier of the termination of the inertial range. There are several reasonable options to choose from.
For example, in the solar wind, the \emph{spectral break scale}, $f_b$, the point at which the power spectrum of the inertial range steepens, is thought to be a good indicator of the onset of dissipation \citep{Leamon_1998, yang2022pressure}. Additionally, the \emph{ion inertial length}, $\di$, and also the ion gyroradius, are frequently found to be in proximity to the break scale, motivating their use as indicators of the onset of the kinetic range \citep{Chen2014, FranciApJ16, Wang2018, Woodham2018, Parashar2019, Cuesta2022_intermittent, LotzEA23}. We note that $\di$ has the advantage that it \change[]{can usually be estimated even when $f_b$ cannot, for example because the time cadence of the data is insufficient to
resolve kinetic (or spectral steepening) scales. In this work we employ both $f_b$ and $ \di $ as approximations for the inner scale. There is, however, an unresolved issue with this procedure of using $f_b$ or $\di$ as the inner scale and this motivates introducing our second approach. Consider the three}{only requires ion density to calculate, rather than the high-resolution magnetic field data needed to resolve spectral-steepening scales and calculate $f_b$. Its disadvantage is that it does not capture the size of the turbulence amplitudes. 

For example, consider the two} different intervals in Fig.~\ref{fig:shifted_spectra}, each with \change[]{the same $\di$ and the same outer scale $\lc$,}{very similar $\di$ and outer scales $\lc$} but with different turbulence amplitudes. The use of $\di$ \remove[]{or $f_b$} as an inner scale in Eq.~\eqref{eq:re_kolmogorov} \change[]{would yield the same $Re$ for all three cases and therefore would}{consequently yields very similar $Re_{\di}$ for both cases because it does} not capture the different dynamics induced by the varying turbulence strengths. 

Fortunately, there is a length scale that typically does depend on the energy of the turbulent fluctuations due to their effect on the shape of the power spectrum. This is the Taylor microscale $\ts$ \citep{Taylor35, Batchelor1970, Matthaeus2008}, hereafter referred to simply as the Taylor scale. See Fig. \ref{fig:shifted_spectra} for one such example.
By employing it in a further reformulation of $Re$ we can capture this strength-of-the-turbulence aspect. The Taylor scale has multiple definitions and can be estimated in several different ways which differ by order unity factors (denoted below by $\gamma$).
These can all be written, for the velocity field $\vv$, as
\begin{equation}\label{eq:taylor_curl}
    \ts^2 = \gamma \frac{\avg{  \vv^2} }
                    {\avg{  (\nabla \times \vv)^2}  }
  ,
\end{equation}
where the value of $\gamma$ 
depends on the specific definition of $\ts$ employed. 
For example, the traditional hydrodynamics usage is that $\ts$ is the curvature (at the origin) of the longitudinal autocorrelation function so that $\gamma = 5$ \citep[e.g.,][p.~211]{Batchelor1970, tennekes1972}. Herein we employ $\gamma = 3$ because it corresponds to the curvature of the \textit{traced} correlation function, which is relatively simple to calculate using spacecraft time series data; see Eqs.~\eqref{eq:autocorr} and \eqref{eq:taylor_expansion}.  

The inertial range comprises the scales $\ell$ which satisfy $ \eta \ll \ell \ll \lc $. 

Moreover, 
in hydrodynamics $\ts$ lies between $\lc$ and the Kolmogorov scale 
  \citep[e.g.,][]{pope_2000}.  
 Eq.~\eqref{eq:taylor_curl}
makes it clear that $\ts$ 
is related to the mean square spatial derivatives of the turbulent flow. 
It can also be interpreted as the ``single-wavenumber equivalent dissipation scale'' \citep{hinze1975}. In plasma systems, the Taylor scale represents small-scale turbulence physics that is not yet well understood, including its relationship to other plasma parameters and the correlation length.


   Re-expressed in terms of the Taylor scale, the 
   exact 
   hydrodynamic
   viscous dissipation rate is
    $\epsilon = \nu\gamma \avg{  \vv^2 }  / \ts^2 $.
   Equating this to the $\epsilon_\text{phenom}$ relation, Eq.~\eqref{eq:eps-phenom}, yields another form for the Reynolds number
(\citet[p.~118]{Batchelor1970}; 
 \citet[p.~67]{tennekes1972}):
\begin{equation}    
\label{eq:re_taylor}
    Re
    \equiv Re_{\ts} 
    = \frac{\gamma}{\Ceps} 
              \left( \frac{\lc}{\ts} \right)^2 ,
\end{equation}
The ratio of Taylor scale to the spectral break scale has been shown to have a direct correlation with the decay rate \citep{Matthaeus2008}. Hence, one would expect this definition of $Re$ to show variation with changing turbulence amplitude and decay rates (as can be seen by the very different values of $Re_{\ts}$ in Fig. \ref{fig:shifted_spectra}).

Note that $\Ceps$ is significantly less than unity. 
  Hydrodynamic simulations and experiments 
    \citep{Sreenivasan98,PearsonEA04}
  indicate that
  \begin{align}
     \Ceps 
     & \approx 
     0.5 \frac{2}{9\sqrt{3}} 
     \approx \frac{1}{15.6} 
   ,
   \label{eq:Ceps-value}
  \end{align}
where in the middle term the 0.5 value is empirical and the other values are associated with ``unit conversion'' from a variant of Eq.~\eqref{eq:eps-phenom} commonly used in the hydrodynamic literature, namely 
  $\epsilon_\text{phenom} = A u_1^3 / \ell_f $;
  here $U^2 = 3u_1^2$ and $\ell_f = 3\lc/2$ is the correlation length for the longitudinal velocity correlation function, all assuming isotropy (see, e.g.,
      \cite{Batchelor1970,tennekes1972,PearsonEA04}).
Thus, in hydrodynamics, with $\gamma = 3$, 
the prefactor 
in Eq.~\eqref{eq:re_taylor} is 
    $\gamma/\Ceps \approx 50$, 
and in Eq.~\eqref{eq:re_kolmogorov} it is 
    $\Ceps^{-1/3}\approx 3$. 
The values in MHD, for solar wind-like conditions, are \change{approximately the same}{$\gamma/\Ceps \approx 27$ and $\Ceps^{-1/3}\approx 2$} (see Appendix~\ref{app:Ceps}). 
These are the values we use in the data analysis reported on below.
However, one should keep in mind that these values pertain to collisional MHD fluid models. The solar wind is an almost collisionless plasma that can, in some circumstances, be well approximated as an MHD fluid.



  \remove{Much of the above discussion has focused primarily on situations associated with Navier--Stokes fluids.  Our main interest herein, however, is the solar wind, which is a nearly collisionless plasma that can, in some circumstances, be well approximated as an MHD fluid.} 
For a system like the solar wind, most velocity measurements have a time cadence 
that is significantly longer 
than kinetic   time  scales (with the exception of measurements from the MMS mission). Because of this, one cannot reliably compute $\ts$ for the velocity field. On the other hand, magnetic field measurements have a significantly higher time cadence, allowing one to explore 
  kinetic 
scale physics. 
Hence most studies in the solar wind compute the Taylor scale for the magnetic field.
Given these constraints, we also work (primarily) with magnetic field data in this study and compute several types of \emph{effective} Reynolds numbers.

A history of estimating magnetic $Re$ in the solar wind is provided in the introduction to \cite{CartagenaSanchez2022}. Prior estimations have used Eq.~\eqref{eq:re_taylor} and applied it to multi-spacecraft measurements, beginning with \cite{Matthaeus2005} and continuing with \cite{Weygand2007, Weygand2009, Weygand2011} and \cite{Zhou2020}. 
Note that these studies 
use $ \gamma / \Ceps = 1 $, 
and thus essentially ignore this prefactor. 
The average values of $\lc$, $\ts$, and $Re_{\ts}$ from these studies are summarised in Table~\ref{table:re_literature}, where we also indicate an appropriate value of $\gamma / \Ceps $ to be used for comparison with the results we obtain herein. 
All these studies used data from a combination of spacecraft at 1\,au, including ACE, Wind, and Cluster, 
and most investigated the relationship between $Re$ and variables such as magnetic field orientation, wind speed, and solar activity. Going beyond 1\,au, this formulation has also been used to estimate $Re$ at Mars \citep{Cheng_2022}, and Voyager data has been used to calculate it at very large distances from the Sun \citep{Parashar2019}. 
Voyager data lacks sufficient resolution to calculate $\ts$ 
and thus $\di$ was used in the Eq.~\eqref{eq:re_kolmogorov} formulation to estimate $Re$. \cite{Cuesta2022_intermittent} supplemented this work with data from Parker Solar Probe and Helios in a survey of variation in $Re$ throughout the heliosphere. 

It is clear from the studies cited above that the Reynolds number plays a pivotal role in understanding solar wind turbulence.  Accurate estimation of $Re$ can be used to validate theoretical predictions such as the enhanced intermittency with increasing $Re$ \citep[e.g.,][]{VanAttaPFL80, Parashar2015, Parashar2019, Cuesta2022_intermittent}, or its correlation with solar activity \citep{Zhou2020, Cheng_2022}. Different formulations need to be compared to bolster these conclusions further. Additionally, a firmer estimate of $Re$ will help refine the minimum scale separation required by an experiment or simulation to faithfully capture the dynamics of such high $Re$ astrophysical systems; this is the so-called ``minimum state'' \citep{Zhou2007, ZHOU_2017}. Therefore, to obtain reliable estimates of the solar wind's (effective) Reynolds number, a thorough comparison of computational techniques and their implications is necessary.

This is the purpose of the present study. A large dataset of measurements from the Wind spacecraft is compiled, allowing us to calculate $Re$ for nearly two decades of data in three different ways: using either $f_b$ (obtained from the magnetic energy spectrum) or $\di$ in Eq.~\eqref{eq:re_kolmogorov}, and using $\ts$, obtained from the autocorrelation function for $\vb$, in Eq.~\eqref{eq:re_taylor}. 

The structure of this paper is as follows. The dataset and its initial cleaning are described in Sect. \ref{sec:data}.  Sect. \ref{sec:method} provides the methods for estimating each of the scales; we calculate $Re_{\ts}$ after first applying the correction to $\ts$ developed by \citet{Chuychai2014}. In Sect. \ref{sec:results}, the three estimators are compared to each other and to the values obtained by the aforementioned studies. Implications and limitations of these results are discussed in Sect. \ref{sec:concl}.

    \section{Data} \label{sec:data}

We use roughly 18 years (2004--2022) of data from NASA's Wind spacecraft to estimate $Re$ at 1\,au. We process $\approx$ 12,000 12-hour intervals in the solar wind. High-resolution (0.092\,s) vector magnetic field data were obtained from the Magnetic Field Investigation (MFI) \citep{Lepping1995}. Wind was launched in 1994 and has operated at the Lagrangian point 1 (L1) since June 2004 in order to study plasma processes occurring in the near-Earth solar wind. This mission has significantly contributed to understanding many aspects of the solar wind, including electromagnetic turbulence \citep{Wilson2020}. 

After downloading the data from NASA/GSFC's Space Physics Data Facility (SPDF), we split it into 12-hour intervals. This interval size is large enough to contain a few correlation lengths but small enough to not average over large-scale variations. \citet{Isaacs2015} demonstrated that (1\,au) intervals of 10-20 hours have ``special significance'' as they represent a range where sufficient correlation times are sampled, making single-spacecraft results coincide with those of multiple spacecraft.

Data gaps are linearly interpolated unless they comprise more than 10\% of the interval, in which case the interval is discarded. (This affected about 4\% of the intervals.) We initially processed 28 years of data, from 1995-01-01 to 2022-12-31, to compute various average quantities as well as turbulence parameters such as the spectral slopes in the inertial and kinetic ranges, rms amplitudes of the magnetic field and velocity, the Taylor scale, and the correlation scale. The complete dataset comprises all available magnetic field data, i.e., intervals containing shocks or from within the Earth's magnetosphere are 
  \emph{not}
removed. However, our analysis in the subsequent sections of this paper focuses only on data from June 2004 and later, a period when Wind was positioned at L1, away from the magnetosphere.

\add{Given that it is also of interest to future analysis how quantities like the Taylor scale relate to other properties of the turbulent plasma system --- such as electron density, cross-helicity, and solar activity --- measurements of electron and proton properties from Wind's 3D Plasma (3DP) instrument were also obtained, along with sunspot numbers from the World Data Center SILSO.}

We note that the ion density from Wind has periods of anomalously small values for a few months. To avoid issues associated with this we therefore always use the electron density as a proxy 
  for the proton density
when calculating all ion inertial lengths, ion plasma betas, and Alfv\'en speeds. Across the 28 years of data, we obtained between 18,000 and 20,000 points for each variable, depending on the amount of missing data. A full list of the variables in the processed (and publicly available) data set can be found in Appendix~\ref{app:dataset}.

    \section{Method} \label{sec:method}
    
\begin{figure} 
    \centering
    \includegraphics[width=\columnwidth]{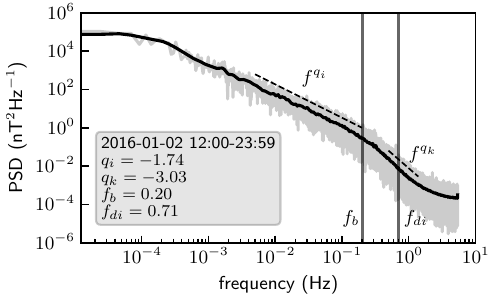}
    \caption{Power spectral density (PSD) of a solar wind magnetic field interval, raw (grey) and smoothed (black). Dashed power-law fits to estimates of the inertial and kinetic ranges return spectral indices $q_i$ and $q_k$. The left-hand vertical line indicates the intersection of these two fits, denoted as the spectral break $f_b$; the ion inertial frequency $f_{\di}$ is indicated by the right-hand vertical line.\remove[]{ For this interval, comprising the second half of 2016-01-02, we obtain $q_i=-1.7$, $q_k=-3.0$, $f_b=0.2$ and $f_{\di}=0.7$.}}
    \label{fig:psd}
\end{figure}

We begin the analysis by determining several slopes for each of the magnetic power frequency spectra obtained from the 12-hour intervals.
Specifically, we perform power-law fits in the inertial and kinetic ranges, denoting the power-law exponents as $q_i$ and $q_k$, respectively. Nominal frequency intervals for the inertial range (0.005--0.2\,Hz) and kinetic range (0.5--1.4\,Hz) were chosen, consistent with those used by \citet{Wang2018}. We then identify the frequency at which (the extrapolation of) these powerlaws intersect, calling this the spectral break frequency $f_b$. An example is shown in Fig.~\ref{fig:psd}. Any outliers, mostly in the form of anomalously large values of $q_k$, are not included in the subsequent analysis, as described in Sect.~\ref{sec:results}. In the following, we will use the time scale associated with the break frequency, i.e., $ t_b = 1/(2\pi f_b) $, as a proxy for the inner (time) scale.

Estimates for the Taylor scale $\ts$ and the correlation scale $\lc$ are also needed and these are both computed using the autocorrelation functions (see Fig.~26 in \cite{Bruno2013}). The (normalized) temporal autocorrelation of the magnetic field fluctuations is given by 
\begin{equation}\label{eq:autocorr}
    R(\tau) 
  = 
    \frac{ \avg{  \vb(t)\cdot \vb(t+\tau)} }
         { \avg{  \vb^2 } } ,
\end{equation}
where $ \vb(t) = \vB(t) - \avg{  \vB(t) }  $ is the magnetic field fluctuation at time $t$. The angle brackets denote a suitable time ensemble average, implemented as a time average in this study. Using Taylor's frozen-in-flow hypothesis, we can convert time separations $\tau$ into length separations $r$. (See Sect.~\ref{sec:concl} for a discussion of the limitations of this hypothesis.) 

Measurement of $\lc$ requires a computation of the autocorrelation function out to very large lags. On the other hand, measurement of $\ts$ requires iterative fitting at very small lags. It would quickly become computationally expensive to use the high-time-cadence data to obtain both quantities. Hence, for each 12-hour interval, the correlation length $\lc$ is computed from a down-sampled low-resolution (5\,s) magnetic field time-series out to roughly 10,000\,s. We use the high-time-cadence (0.092\,s) magnetic field data to compute autocorrelation functions only up to a lag of 9.2\,s; this is used to compute the magnetic Taylor scale $\ts$.

\begin{figure} 
    \centering
    \includegraphics[
    width=\columnwidth,
    ]{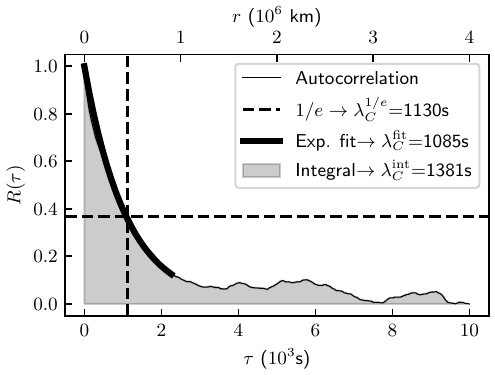}
    \caption{A demonstration of the three methods used to calculate $\lc$ using an interval comprising the second half of 2016-01-02. These include the 1/e (``e-folding'') method, giving $\lc^{1/\e}$; the exponential fit method, giving $\lc^\text{fit}$; and the integral method, giving $\lc^\text{int}$.}
    \label{fig:corr_scale_methods}
\end{figure}

The correlation scale $\lc$ for $\vb$ can be estimated from $R( \tau ) $ 
in three different ways, as shown in Fig.~\ref{fig:corr_scale_methods}.  
We can perform an exponential fit, we can find the separation at which the function falls to 1/e, or we can take the integral of the function (\(\lc = \int^{\infty}_{0} R(\tau) \, \d\tau\)). The exponential fit method is frequently used in the literature \citep{Matthaeus2005, Zhou2020, Bandyopadhyay2020, Phillips2022}; multiple exponential fits and a third-order polynomial have also been used \citep{Weygand2009, Weygand2011, Cheng_2022}. In any case, this requires a decision about how much of the autocorrelation to fit to. In this work, we fit a single exponential to a range that extends to twice the value of the correlation scale as obtained by the $1/\e$ method. We compute $\lc$ from the low-resolution autocorrelation using each of these three methods to evaluate their consistency.

While it is straightforward to compute the Taylor scale in simulations, where one has access to the full three-dimensional information, when working with time series data from experiments we need to resort to an approximation. 
Since $\ts$ can be defined as the radius of curvature of the autocorrelation function at the origin, we may use this definition to estimate it. 
(We do not yet need to convert to spatial lags, so we work with the time-domain equivalent, $\tau_\text{TS}$). 
This follows from the Taylor expansion of the autocorrelation for 
  $\tau \rightarrow 0 $ 
  \citep{Batchelor1970, tennekes1972}:
\begin{equation}\label{eq:taylor_expansion}
    R(\tau) = 1 - \frac{\tau^2} {2\tau_\text{TS}^2}+...
\end{equation}

\begin{figure} 
    \centering
    \includegraphics[
    width=0.95\columnwidth,
    ]{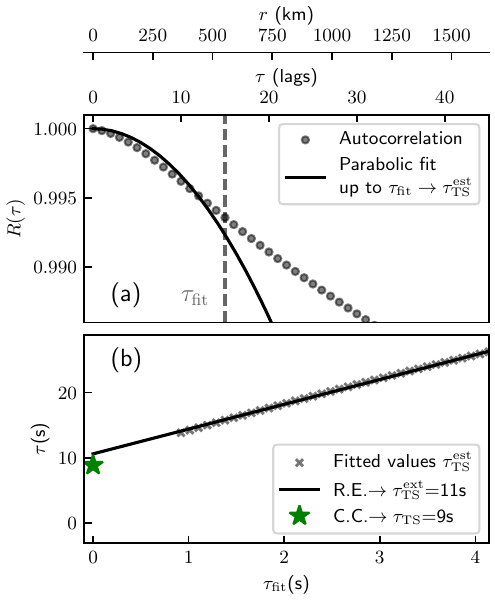}
    \caption{An example of the process of refining the estimate of the Taylor scale $\tau_\text{TS}$, using an interval comprising the second half of 2016-01-02. The three horizontal scales show the separations in units of lag, time, and (Taylor frozen-flow equivalent) distance. 
    \textbf{Panel (a)}: Firstly, a parabola is fit to the autocorrelation from the origin up to various  
    values of $\tau_\text{fit}$. 
    In this example the fit is for lags less than $\tau_\text{fit} = 15 $. 
    The $x$-intercept of each parabola (which is off-scale for this plot) produces an initial estimate $\tau_\text{TS}^\text{est}$. 
    \textbf{Panel (b)}: Next, each of these estimates is plotted against $\tau_\text{fit}$. A straight line is fit to these points and extrapolated back to $\tau=0$, returning the Richardson extrapolation (R.E.) estimate, $\tau_\text{TS}^\text{ext}$. Finally, the Chuychai correction (C.C.) is applied using the kinetic range slope ($ q_k = -3.03$ in this case) in Eq.~\eqref{eq:chuychai_correction}. This yields our final estimate, $\tau_\text{TS}$; 
    we obtain a value of 9\,s, 
    or approximately 4,000\,km, 
    for this particular interval.
    }
    \label{fig:taylor_scale_method}
\end{figure}
In practice, this means fitting a parabola to $R(\tau)$ at the origin and  requires the high-resolution data provided by Wind so that we have enough observations at small separations. It also requires an important decision: how much of this high-resolution autocorrelation do we fit to? (Larger ranges result in systematically larger estimates.) In order to reduce the subjectivity of this decision, the Richardson extrapolation technique was introduced in this context by \citet{Weygand2007}: by fitting to a range of values of maximum lag $\tau_\text{fit}$, then extrapolating back to 0 lags, we obtain a refined estimate, $\tau_\text{TS}^\text{ext}$. In the aforementioned work, the authors showed an apparent convergence of the final estimate given by this technique as $\tau_\text{fit}$ increases. However, \citet{Chuychai2014} showed with simulated data that, in fact, this convergence depends upon the slope of the power spectrum at high frequencies. In light of this, they produced a multiplicative correction factor, $ r(|q|)$, 
that is a function of this slope, given as 
\begin{equation}\label{eq:chuychai_correction}
r(|q|) = 
\begin{cases} 
  -0.64(\frac{1}{|q_k|})+0.72, & \text{when } |q_k| < 2 \\
  -2.61(\frac{1}{|q_k|})+1.70, & \text{when } 2 \leq |q_k| < 4.5 \\
  -0.16(\frac{1}{|q_k|})+1.16, & \text{when } |q_k| \geq 4.5
  .
\end{cases}
\end{equation}
We also apply this correction to our estimates, with the procedure we follow depicted in Fig.~\ref{fig:taylor_scale_method}. This gives us a final estimate $\tau_\text{TS}$. We fit from a minimum lag of 1, equal to the time cadence (0.092\,s), up to a maximum lag $\tau_\text{fit}$ which was varied between 10 and 50 lags.

Finally, using the various (magnetic) scales, determined as outlined above, we calculate estimates for effective Reynolds numbers in three distinct ways.  Specifically, we use 
$ \lc^\text{fit} $  (or its time scale analog) as the outer scale and
\begin{enumerate} \itemsep=-0.4ex 
 \item[(i)] Eq.~\eqref{eq:re_kolmogorov} with $\di$ as the inner scale and \change[]{$\Ceps^{-1/3} = 3$}{$\Ceps^{-1/3} = 2$}; 
 \item[(ii)] Eq.~\eqref{eq:re_kolmogorov} with $t_b = 1/(2\pi f_b)$ as the inner (time)scale and \change[]{$\Ceps^{-1/3} = 3$}{$\Ceps^{-1/3} = 2$}; 
 \item[(iii)] Eq.~\eqref{eq:re_taylor} with \change[]{$\gamma/\Ceps = 50$}{$\gamma/\Ceps = 27$}.
\end{enumerate}

    \section{Results} \label{sec:results}

Our analysis uses data from the period June 2004 to \change{October}{December} 2022 when Wind was always situated in the solar wind at L1. 
In about 6\% of the intervals the slope of the kinetic range, $ q_k $, was unusually shallow
(meaning $ |q_k| < 1.7 $) and therefore the final (corrected) estimate of the Taylor scale came out to be negative. These outlier intervals were removed from the following analysis but will be investigated in future work. 

    \subsection{Correlation scale}

\begin{table}  
\centering
\begin{tabular}{||c c c c||} 
 \hline
 Method & Mean (km) & Median (km) & SE (km) \\ [0.5ex] 
 \hline\hline
 Exponential fit & 899,000 & 769,000 & 5,000 \\ 
 $1/\e$ & 942,000 & 797,000 & 5,000\\
 Integral & 880,000 & 808,000 & 4,000 \\ [1ex] 
 \hline
\end{tabular}
\caption{Statistical summary of estimates of the magnetic field correlation length $\lc$ by different methods. \add{The standard error (SE) gives the expected variation of the mean between samples of this size.}}
\label{table:corr_scale}
\end{table}

\begin{figure*} 
    \centering
    \includegraphics[
    width=\textwidth]{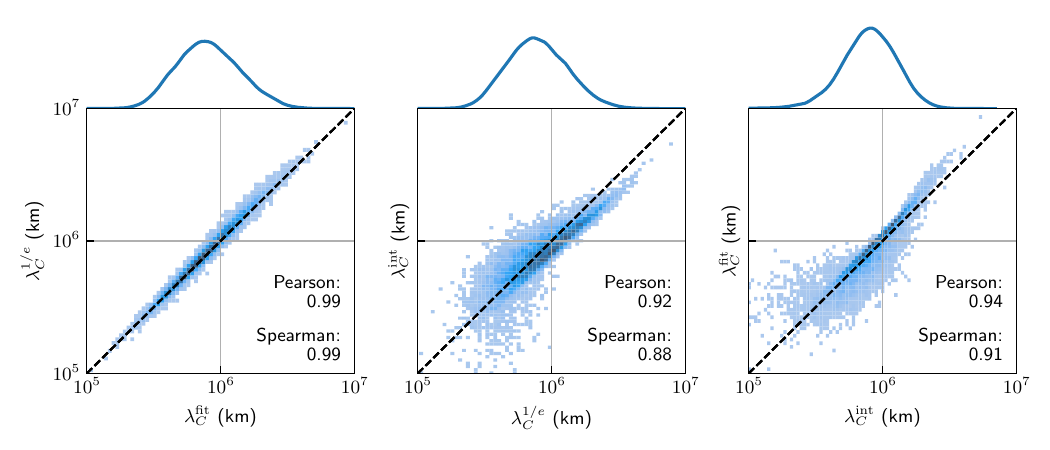}
        \caption{Joint (2D) histograms of the three $\lc$ estimates with Pearson (linear) correlation and Spearman (rank) correlation values and a dashed line of equality.  Marginal (1D) histograms of the $x$-variable are shown above each plot. The $x$- and $y$-axis limits have been set so as to include the bulk of the data but exclude outliers. $\lc^\text{fit}$: exponential fit method,  $\lc^{1/\e}$: 1/e method,  $\lc^\text{int}$: integral method.}
    \label{fig:corr_scale}
\end{figure*}
Table~\ref{table:corr_scale} gives summary statistics of each of the three estimates of the correlation length of the magnetic field, $\lc$, and Fig.~\ref{fig:corr_scale} shows their marginal and joint distributions. Given the wide distribution of values, all values are in line with those previously reported in the literature at 1\,au, i.e., approximately $10^6$\,km (see Table~\ref{table:re_literature}). Noting the logarithmic scaling of the axes in this figure, we qualitatively find that the probability distribution function of each estimator is log-normal. This is consistent with the results of \citet{Ruiz2014} as well as the distribution of many other solar wind quantities such as proton temperature, plasma beta, and Alfv\'en speed \citep[e.g.,][]{hundhausen1970solar, Burlaga2000,mullan2006,veselovsky2010algebra}. In particular, the correlation scales are positively skewed, with means larger than the corresponding medians. Looking at the joint distributions, we see that the exponential fit and 1/e methods agree very well with each other, with very high values of 0.99 for both the Pearson and Spearman correlations, and most of the points lying close to the equality line. (The Spearman correlation uses ranks to measure the monotonicity of the relationship between two variables, rather than measure their linear association.) This agreement is not surprising given the large-scale statistical homogeneity of the solar wind. The autocorrelation functions typically show 
approximately exponential fall-off (see, e.g., Fig.~\ref{fig:corr_scale_methods}), with deviations from $ \lc^{1/\e} \approx \lc^\text{fit} $ only occurring for intervals that do not show steady turbulence \citep{Ruiz2014}. 

In contrast, the integral scale $\lc^\text{int}$ shows a moderate degree of scatter against either of the other two estimates, with correlations of between 0.88 and 0.94. The greatest degree of scatter is present for values of $\lc^\text{int}$ less than about $ 10^6 $\,km. 
This disagreement is likely due to occasional numerical issues with calculating the integral of the autocorrelation. Ideally, the integral is computed out to infinity as $R$ asymptotically decays to 0. However, the finite size of the intervals and the slight departures from ``textbook-like'' homogeneity and isotropy in some intervals could introduce discrepancies between this and the exponential estimates. Nonetheless, we conclude that, to a reasonable approximation, all three methods give equivalent estimates for $\lc$.

    \subsection{Taylor scale}

\begin{figure}  
    \centering
    \includegraphics[width=\columnwidth]{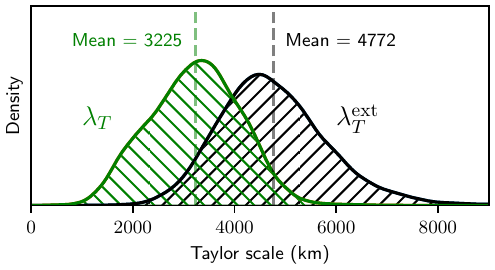}
        \caption{Distributions of the uncorrected ($\ts^\text{ext}$) and corrected ($\ts$) versions of the Taylor scale. Dashed vertical lines indicate mean values for each distribution.  The $x$-axis limits have been set so as to include the bulk of the data but exclude outliers.
        }
    \label{fig:taylor_scale_hists}
\end{figure}

\begin{figure*}[btp]  
    \centering
    \includegraphics[width=\textwidth]{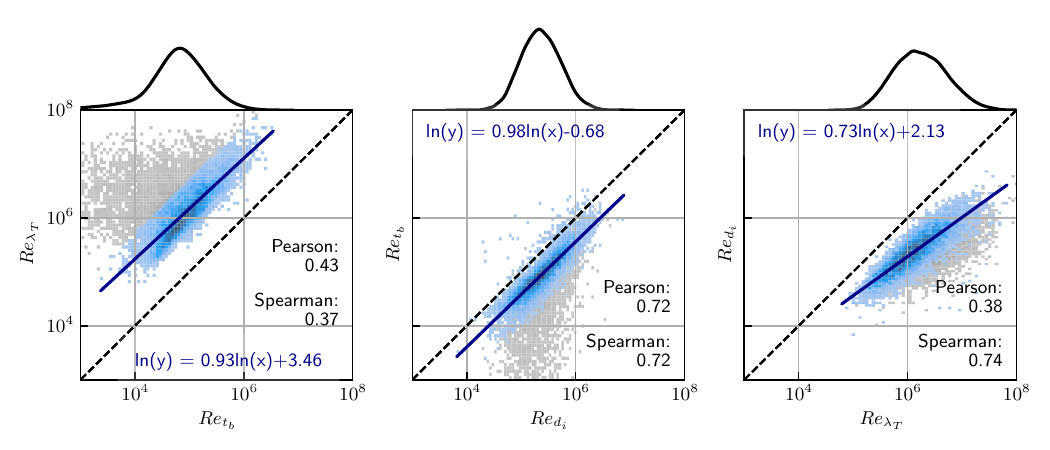}
        \caption{Joint (2D) histograms of the three $Re$ estimates with Pearson correlation and Spearman (rank) correlation values, log-space regression line fits, and a dashed line of equality; and (above top axis) marginal (1D) histograms of the $x$-variable in each plot. The $x$- and $y$-axis limits have been set so as to include the bulk of the data but exclude particularly extreme outliers beyond these limits. Remaining outliers are shaded grey.  Correlation coefficients and marginal histograms are for all data values, whereas regression lines are fitted to only the majority subsets of the data shown in blue (see text).}
    \label{fig:re_panels}
\end{figure*}

Fig.~\ref{fig:taylor_scale_hists} shows marginal distributions of both the uncorrected and corrected versions of $\ts$  for the magnetic field. Both have quasi-Gaussian distributions, with a few large outliers. The distribution of $\ts$ computed after applying the Chuychai correction factor is shifted to the left because the (multiplicative) correction factor is almost always less than 1, except for the 1\% of intervals with particularly steep kinetic range slopes ($q_k <-3.7$). The mean $q_k$ is $-2.64$, resulting in an average correction factor of $-2.61/2.64+1.7=0.71$, following Eq.~\eqref{eq:chuychai_correction}. We therefore end up with a mean of $\ts$ that is about two thirds that of $\ts^\text{ext}$. We find that this final mean of \change{3,110}{3,225}\,km is in good agreement with the literature (see Table~\ref{table:re_literature}). Prior estimates of $\ts$ in the solar wind at 1\,au vary between $\sim 1,000$\,km and $\sim 7,000$\,km, values which lie within the distributions of $\ts$ (both extrapolated or Chuychai-corrected) shown in Fig.~\ref{fig:taylor_scale_hists}. 

\begin{table*}
\centering
\begin{tabular}{||l l l l r l||} 
 \hline
 Authors (Year) & Spacecraft & $\lc$ ($10^6$\,km) & $\lambda_{T}$ (km) & $Re_{\ts}$ & \\ [0.5ex] 
 \hline\hline
 \citet{Matthaeus2005} & ACE-Wind-Cluster & 1.2 & 2,478 $\pm$ 702 & 230,000 & ($\times$27)\\
 \citet{Weygand2007} & Cluster & 1.2 (from above) & 2,400 $\pm$ 100 & 260,000 & ($\times$27)\\
 \citet{Weygand2009} & ACE-Wind-Cluster + 6 others & 2.92 & 1,000 $\pm$ 200 & 12,600,000 & ($\times$27)\\
 \citet{Weygand2011} & ACE-Wind-Cluster + 8 others & 1--2.8 & 1,200--3,500 & 4,000,000 & ($\times$27)\\
 \citet{Zhou2020} & ACE-Wind-Cluster & 1.14 & 2,459 & 300,000* & ($\times$27)\\
 \citet{Bandyopadhyay2020} & MMS & 0.32 & 6,933 & 2,000** & ($\times$27)\\
 \textbf{This work} & \textbf{Wind} & \textbf{\change{0.871}{0.899}} & \textbf{\change{3,110}{3,220}} & \textbf{\change{6,330,000}{3,406,000}} & \\ [1ex] 
 \hline
\end{tabular}
\caption{Average estimates of $\lc$, $\ts$ and an effective Reynolds number in the solar wind at 1\,au, with the latter calculated using Eq.~\eqref{eq:re_taylor}. Shown are the values determined in this work (given in bold) and in some previous studies.  Note that for direct comparison with this work, the $Re_{\ts}$ values from these earlier studies should be multiplied by the previously neglected prefactor of $\gamma/\Ceps = 27$, as indicated by the `$(\times 27)$' in the final column. This factor is already included in our estimates. 
When calculating $ \ts $ all studies listed employed $ \gamma=3 $, sometimes without explicitly stating so.
All studies used at least one exponential fit to compute $\lc$. All except \citet{Matthaeus2005} and \citet{Bandyopadhyay2020} used Richardson extrapolation to compute $\ts$; none, other than this work, used the Chuychai correction. Values are expressed as ranges when the study grouped scales by other variables such as magnetic field orientation.
\\ * This mean value was reported in a follow-up article \citep{GZhou2021_ApJ}.
\\ ** $Re$ was not calculated explicitly in this article.}
\label{table:re_literature}
\end{table*}

    \subsection{Reynolds number}
    
Having obtained estimates for the correlation length and Taylor scale of the magnetic field fluctuations (and also for $\di$ and $\tb$) we may use the procedures detailed at the end of Sect.~\ref{sec:method} to calculate three distinct effective Reynolds numbers. Fig.~\ref{fig:re_panels} shows the marginal and joint distributions of these different estimates, as well as regression line fits. After applying a logarithmic transformation each distribution appears approximately Gaussian, suggesting a log-normal distribution. Comparing these marginal distributions and the summary statistics given in Table~\ref{table:re_stats}, we can see that the three estimates span multiple orders of magnitude, with, very roughly, $Re_{\ts} \sim 10 Re_{\di} \sim 30 Re_{\tb} $, for 
the mean 
values.

The joint distributions show considerably more scatter than those of the $\lc$ estimates. The strongest linear association between any two estimates is that between $Re_{\di}$ and $Re_{\tb}$ (Pearson correlation = \change{0.75}{0.72}). This is shown by the majority of points lying in a relatively thin linear band \add{close to the equality line}. We can also see that the $Re_{\tb}$ estimates tend to be smaller than $Re_{\di}$. This is an indication that the break scale is typically larger than $\di$ by a factor of 2-3 in the solar wind \citep{Leamon_1998}. A dependence of the break scale on plasma $\beta$ is also well known \citep{Chen2014,FranciApJ16}. The statistical details of any such potential correlations will be explored in a follow-up study.

$Re_{\ts}$ shows a much weaker linear association with the other two methods of only \change{0.44}{0.38} (with $Re_{\di}$) and 0.43 (with $Re_{\tb}$). In addition to having the lowest Pearson correlation, $Re_{\ts}$ and $Re_{\tb}$ also have the lowest Spearman correlation, showing that even after accounting for outliers, which have less influence on this latter metric, it still remains a rather weakly positive association. On the other hand, outliers do have a clear influence on the linear association of $Re_{\ts}$ vs.\ $Re_{\di}$, shown by the substantial increase in the Spearman correlation (\change{0.69}{0.74}) over the Pearson correlation (\change{0.44}{0.38}).

Despite these only moderately strong associations between the estimates, it is important however to note the density of the points. All these distributions show significant scatter of a small population in which the estimates differ by up to an order of magnitude. Notably, the joint distributions of $Re_{\tb}$ have a roughly triangular sub-population of points that shows little to no relationship with the other estimates.  This is seen in the upper left of the plot of $Re_{\ts}$ vs.\ $Re_{\tb}$, and the lower right of $Re_{\tb}$ vs.\ $Re_{\di}$. This population (identified as $Re_{\ts}/Re_{\tb} > 50$) represents about 27\% of all observations, and is shown as the grey points in Fig.~\ref{fig:re_panels}. After removing this population, all correlations increase to at least \change{0.8}{0.68}. The potential reasons for significantly larger $t_b$ and hence a smaller $Re_{\tb}$ could include errors in automated fitting and extreme intervals with atypical 
 power spectra. 
As with the other outliers, a detailed investigation of these is deferred to a follow-up study. 
In cases 
where the power spectrum is well-behaved, with typical inertial and kinetic range slopes ($q_i$ and $q_k$) and a well-defined breakpoint, it might be safe to estimate $Re_{tb}$ and multiply it by 30 to estimate $Re_{\ts}$.


\begin{table}  
\centering
\begin{tabular}{cccc} 
 \hline
 $Re$ & Mean  & Median  & \change{SD}{SE} \\ [0.5ex] 
 \hline
 $Re_{\tb}$ & \change{173,000}{116,000} & \change{95,000}{64,000} & \change{263,000}{2,000} \\ 
 $Re_{\di}$ & \change{521,000}{330,000} & \change{356,000}{226,000} & \change{545,000}{3,000}\\
 $Re_{\ts}$ & \change{6,330,000}{3,406,000} & \change{3,130,000}{1,686,000} & \change{14,200,000}{68,000} \\ [1ex] 
 \hline
\end{tabular}
\caption{Statistical summary of the estimates of effective $Re$ obtained by the different methods. (\change{SD = standard deviation}{SE = standard error of the mean}.)}
\label{table:re_stats}
\end{table}


As well as the agreement between methods, it is also of interest how our estimates of $Re$ match up with those previously reported. In particular, given the prevalence of the Taylor scale method, we compare values of $Re_{\lambda_T}$ in Table~\ref{table:re_literature}. The values for $Re$ in this table vary by a factor of $\approx6000$, from \change{100,000}{54,000} to \change{631,000,000}{340,000,000} after multiplying by the prefactor. The mean value of \change{$Re_{\ts}= 6,330,000$}{$Re_{\ts}= 3,406,000$} from the present study is on the same order of magnitude as the results from three of the previous works. The much larger values given in \citet{Weygand2009} and \citet{Weygand2011} were mainly attributed to the smaller values obtained for $\ts$. Conversely, \citet{Bandyopadhyay2020} noted that their value of $\ts$ calculated from a single 5-hour interval of MMS data was about three times larger than previous estimates, while their estimation of $\lc$ was smaller than other estimates. Hence they computed a much smaller value of $Re$. Three reasons were suggested for this: 
1) interval length, separation and mixing effects, 
2) intrinsic variability in the solar wind, and 
3) differences between the 
  geometric formation of the 
Cluster (to which they were comparing their results) and MMS spacecraft. 
Our work herein emphasizes that point 2) is indeed pertinent. 
In particular, our results show the considerable intrinsic variability of the solar wind properties (particularly $\lc$ and $\ts$), giving rise to variability in the values of 
   effective
   $Re$. 
On the plus side, this sampling variability suggests that the results of all the cited studies may in fact be consistent with each other, as they lie within the distribution of values found in our study.

    \section{Conclusion} \label{sec:concl}


We present a thorough investigation and review of calculating estimates of (effective) Reynolds numbers for the solar wind at 1\,au, using 18 years of data from NASA's Wind spacecraft. As this dataset lacks high-time-cadence velocity measurements, we employ magnetic field data to estimate $\lc$ and $\ts$ for the magnetic field. These are assumed to be comparable to their velocity field equivalents,  in line with previously published results. More precisely, in using the magnetic length scales in~\eqref{eq:re_taylor} we are assuming that $ \lc^b / \ts^b \approx \lc^v / \ts^v $.

We first compare three different ways of calculating the correlation scale and find good agreement between all methods, albeit with a greater scatter for the integral method. The mean values obtained for $\lc$, between \change{853,000}{880,000}\,km and \change{913,000}{942,000}\,km, are consistent with previously reported values of about $10^6$\,km.

We then apply the correction factor developed by \citet{Chuychai2014} to our estimates of the Taylor scale in order to reduce any remaining bias after using the Richardson extrapolation technique. This correction factor typically reduces the estimate of the Taylor scale, significantly shifting the distribution to smaller values. In particular, the mean reduces to \change{3,110}{3,225}\,km, roughly 2/3 of the uncorrected mean value of \change{4,634}{4,772}\,km. Both values are consistent with previous estimates, given the wide spread of the distribution.

Finally, we compute effective Reynolds numbers using three distinct methods. It should be noted that we include proportionality factors in our calculations. In particular, we highlight that the factor in $Re_{\ts}$ of \change{$\Ceps \approx 50$}{$\Ceps \approx 27$} was not included in many previously published estimates (see~Table~\ref{table:re_literature} and Eqn.~\eqref{eq:re_taylor}). 

While very strong correlations were observed for the three different methods of estimating $\lc$, the correlations between the associated estimates of the effective Reynolds number were only moderate to strong, with a considerable amount of scatter. The mean values determined by these methods ranged from \change{173,000}{116,000} for $Re_{\tb}$ to \change{6,330,000}{3,406,000} for $Re_{\ts}$. Putting these into perspective, previously reported values of $Re$ at 1\,au exhibit substantial variability, ranging from approximately $10^6$ to $10^8$. Most of our estimates of $Re$ comfortably fit within this distribution, though an outlier population of small values of $Re_{t_b}$ warrants future investigation.

Ultimately, we conclude that more theoretical work is needed to better understand which definition of an effective Reynolds number of the solar wind is most appropriate. The key task is to identify scales that have a physical meaning. For the $t_b$ or $\di$ approximations of the inner scale, the implication is that ion-scale physics plays the most significant role in energy dissipation and terminates the inertial range. This, however, discounts the role of a sub-ion-scale cascade and its implications for electron physics \citep{Matthaeus2008, Alexandrova2009,Sahraoui2009, schekochihin2009astrophysical, boldyrev2013toward}. Moreover, these estimates are insensitive to the variability of the power input at large scales and hence the cascade rate. On the other hand, the $\ts$-based estimate of $Re$ indirectly folds in the cascade rate through its  dependence (empirically in the solar wind but directly in hydrodynamics) \citep{pope_2000, Matthaeus2008}. This makes $Re_{\ts}$ a more physically motivated estimate amongst the three considered.

Moreover, having obtained statistical relationships between different estimates, these can be leveraged in situations where only one estimate is calculable. The decision on which estimator to use rests on the assumptions one elects to make and on the resolution of the available data. These considerations are summarized below.

\begin{itemize}
    \item $Re_{\tb}$ requires calculation of the spectral break scale. This process is subject to varying methods and numerical challenges, including spectra that do not always show clear breaks. In our work, we calculated $t_b$ as the intersection of (the extrapolation of) two power-law fits to magnetic field spectra, which requires decisions on what intervals to choose for the inertial and dissipation ranges.
    
    \item Alternatively, one can simply use the ion inertial length $\di$ to approximate the break scale and calculate $Re_{\di}$ \citep{Parashar2019, Cuesta2022_intermittent}. This requires only the ion density (and correlation length). However, it appears that changing solar wind conditions affect which scale is best associated with the spectral break. Specifically,  $\di$ is the best approximation at low plasma $\beta$ values, the ion gyroradius $\rho_i$ is best at high $\beta$ values \citep{Chen2014}, and for typical solar wind values of $\beta\approx 1$, the ion cyclotron resonance scale is the best \citep{Woodham2018}. Under conditions where one might not have high-time-cadence measurements of the desired variables, it is likely that one could still easily obtain reasonable estimates for both $\di$ and the outer scale (e.g., $\lc$) and employ these to estimate an effective Reynolds number.
    
    \item Using the Taylor scale-based Reynolds number, $Re_{\ts} $, is a more robust formulation for estimating $Re$ than the two listed above because of its empirical dependence on the cascade rate \citep{Matthaeus2008}. 
    This benefit is shown by the prevalence of this formulation in the literature \citep{Matthaeus2005, Weygand2007, Weygand2009, Weygand2011, Zhou2020, Phillips2022}. 
    We show here that the use of a correction factor \citep{Chuychai2014} makes a significant difference in the estimates of $\ts$ and hence $Re_{\ts}$.
    However, as discussed above, the $\ts$ definition of $Re$ has a 
    proportionality factor that is, in part, determined by a phenomenological or empirical fitting for the 
      (kinetic energy) dissipation rate.  
    Calculating $\ts$ also requires high-resolution data, which is not always available;
     for example, outer heliosphere Voyager observations are so restricted \citep{Parashar2019}. 
    Furthermore, while this does not affect the validity of this formulation, we note that weak cascade rates have been shown to result in $\ts$ being smaller than the break scale, inverting the hydrodynamic ordering \citep{Matthaeus2008}. This is believed to be due to greater 
      relevance of 
    electron dissipation 
      (relative to proton)
      in these weak cascade circumstances  
    \citep{matthaeus2016turbulence}.
\end{itemize}

A limitation of this work is that it relies on the Taylor hypothesis to convert from single-spacecraft time separations to length separations. This assumes that the bulk flow is sufficiently fast that local variations in time can be effectively ignored \citep[see][for solar wind context]{Verma2022}. The Taylor hypothesis relates to the well-studied ``sweeping" hypothesis, whereby large-scale fluctuations sweep (i.e., advect) smaller-scale fluctuations \citep{kraichnan1965, tennekes_1975, YeZhou_2021_PR}. Although invoking Taylor's hypothesis at kinetic scales might introduce substantial inaccuracies, it has nonetheless been shown, numerically and from observations, to be a reasonably good approximation up to second-order statistics \citep{Perri_2017, Chhiber2018, Roberts2022}. This is also true under a model that incorporates sweeping phenomenology \citep{Bourouaine_2019, Perez2021}.
Furthermore, for the present analysis, we note that this assumption does not affect the results for $Re_{\ts}$, because both of the scales involved are in fact left as time scales for this calculation. 
Another aspect that we did not address in this study is the issue of anisotropy in $\lc$ and $\ts$ \citep[e.g.,][]{Weygand2009,Weygand2011,Cuesta_2022_isotropy, Roy2022}. We reiterate that no data filtering was conducted, except to remove intervals with significant missing data, limit the analysis time period to June 2004 onward, and remove outliers where $|q_k| < 1.7$. 

Finally, we envisage that the full 28-year dataset and the accompanying code that we have provided as a data product will be useful to the scientific community for future large-scale statistical analysis and data mining, for Wind and other missions. Future work will start investigating correlations, dimensionality reduction, and machine learning models.

    \section{Acknowledgments}
    
\change[]{We acknowledge the use of OMNI and Wind data from NASA/GSFC's Space Physics Data Facility }{We are grateful to the Wind MFI and 3DP instrument teams for the data and NASA GSFC's Space Physics Data Facility for providing access to it} \footnote{\url{https://spdf.gsfc.nasa.gov/}}. We acknowledge the World Data Center SILSO at the Royal Observatory of Belgium in Brussels for providing the sunspot data \footnote{\url{http://www.sidc.be/silso/}}. \add[]{We also thank Bill Matthaeus for providing feedback on the manuscript.}

\section{Author contributions}
\begin{itemize}
    \item TNP: Conceptualized and supervised the project and edited the draft manuscript.
    \item DW: Refined and extended the pipeline, created the final dataset, and wrote the draft manuscript.
    \item SO: Identified and calculated the prefactors in the Reynolds number equations.
    \item KDL: Conducted preliminary analysis with guidance from TNP and MF.
    \item All authors discussed results and implications and helped edit the final manuscript.
\end{itemize}

    \appendix

  \section{Determination of the \texorpdfstring{$\Ceps$}{Cepsilon} prefactor for MHD}
  \label{app:Ceps}

A standard phenomenological estimate for the kinetic energy dissipation rate ($\epsilon^v$) 
in Navier--Stokes turbulence is
\begin{equation}  \label{eq:eps-phenom-Ceps-A}
  \epsilon_\text{phenom}
     = 
   A \frac{ u_1^3 } { \ell_f^v }
     =
   \Ceps \frac{U^3} { \lc^{ } } 
\end{equation}
  \cite[e.g.,][]{Batchelor1970, tennekes1972, pope_2000}.
Here, 
    $ \avg{  \vv \cdot \vv }  = U^2 $
and
 $u_1$ is the rms velocity for one component of $\vv$.
Also    $ \ell_f^v $ 
is the correlation length associated with the longitudinal correlation function
  \citep{Batchelor1970}, 
whereas $\lc$ is that for the traced correlation function 
  $ R^v(r) = \avg{  \vv(\vx) \cdot \vv(\vx + \vct{r}) }  $, 
equivalent to our $\lc$ definition in the main text.
The dimensionless coefficients $A$ and $\Ceps$ are treated as constants that may be determined using experiments and/or simulations
  \citep{Sreenivasan98,PearsonEA04}.  
For isotropic turbulence, the relations 
  $ U^2 = 3 u_1^2 $,
  $ \lc = 2 \ell_f^v / 3 $, and 
  $ \Ceps = 2 A / (9 \sqrt{3}) $ hold.
As the middle `component-based' version is founded on the assumption of isotropy, in this work we instead employ the rightmost `trace-based' variant which does not assume isotropy; this is given as Eq.~\eqref{eq:eps-phenom} above.

In the literature a variety of notations are in use for what we have called $A$ and $\Ceps$, and indeed some works use $\Ceps$ for the $A$ in Eq.~\eqref{eq:eps-phenom-Ceps-A} 
    \citep[e.g.,][]{PearsonEA04};
clearly this should not be confused with the $\Ceps$ we employ herein.
For clarity, and
in line with the notation of \citet[eq.~6.1.1]{Batchelor1970}, 
we always use $A$ to denote a component-based fitting value. 

We wish to determine a value for $\Ceps$ that is applicable in MHD. This requires taking into account the dissipation of magnetic as well as kinetic energy.  
With superscripts $v$ and $b$ denoting velocity and magnetic quantities, respectively, we may write the total energy decay rate as
 $  \epsilon^\text{MHD} 
    = 
  \epsilon^v + \epsilon^b $.

Using an Elsasser variable 
  ($ \vz_\pm = \vv \pm \vb $)
von K\'arm\'an--Howarth equation analysis for incompressible MHD, 
  \cite{LinkmannEA15,LinkmannEA17-Ceps}
developed a theory for
    $ A^\text{MHD} $ 
(denoted $ C_{\epsilon,\infty} $ therein).
For simplicity here we restrict attention to situations with low cross helicity,  i.e., $\avg{\vv \cdot \vb } \approx 0 $.
Consequently,
    $ \avg{|\vz_+|^2} \approx \avg{|\vz_-|^2} = Z^2 = 3W^2 $,
and
the two longitudinal Elsasser correlation lengths, 
  $ \ell_f^\pm $,
are approximately equal.
Assuming further that the longitudinal correlation lengths for $\vv$ and $\vb$ are also approximately equal, the 
  (low cross helicity)
Linkmann et al.\ result is equivalent to
\begin{equation} \label{eq:diss_factor}
  \epsilon^\text{MHD}_\text{phenom} 
    = 
  A   \frac{W^3} {\ell_f^+} 
   =
  \frac{2A} {9\sqrt{3}}
  \frac{Z^3} {\lc^v}   
 , 
\end{equation}
where  $A  \equiv A^\text{MHD} $ 
and we have made use of the isotropic relation
    $ \ell_f^v = \frac{3}{2} \lc^v $.
Eq.~\eqref{eq:diss_factor} 
applies to the total (viscous plus resistive) dissipation and
is the 
MHD analogue of Eq.~\eqref{eq:eps-phenom} above.
Although, formally, it only applies for 
low cross helicity cases, it is likely to be approximately valid under somewhat more general circumstances
  \citep{LinkmannEA17-Ceps,BandyopadhyayEA18-Ceps}.




Next, we make use of the Alfv\'en ratio, 
    $ r_A = U^2/ \avg{\vb^2} $, 
to write $ \epsilon^\text{MHD}$ 
in terms of $\epsilon^v $.   
In terms of $U$ and $r_A$, 
the zero cross helicity form for $Z^2$ is
\begin{equation}
  Z^2
     =  U^2 + \avg{\vb^2}
     =  U^2 \left( 1 + \frac{1}{r_A} \right)
     =  U^2 F(r_A) 
 , 
  \label{eq:zU-rA}
\end{equation}
which defines $ F(r_A) $.

Recall also that 
 $ \epsilon^\text{MHD}
      =
  \nu \avg{ \vct{\omega}^2 } + \mu \avg{ \vj^2 }
  $,
with $\nu$ the kinematic viscosity, 
$\mu$ the magnetic resistivity, 
  $ \vct{\omega} = \nabla \times \vv$ the vorticity,
  and
  $ \vct{j} = \nabla \times \vb $ the electric current density.
Assuming a Prandtl number of order unity 
  ($\nu \approx \mu$) 
and that 
  $ \avg{ \vj^2 } / \avg{ \vct{\omega}^2} \approx  1/r_A $, 
  as is commonly seen in MHD simulations,
 $ \epsilon^\text{MHD} $
can be re-expressed without explicit reference to the dissipation coefficients:
\begin{equation} \label{eq:diss_alfven_ratio}
    \epsilon^\text{MHD}  \approx  \epsilon^v F(r_A)    
  .  
\end{equation}

Finally, using Eqs.~\eqref{eq:diss_factor} and \eqref{eq:diss_alfven_ratio}, 
we can write a (small cross helicity) approximation for the kinetic energy dissipation rate in MHD:
\begin{equation}
   \epsilon^v
      \approx
   \left( \frac{2A} {9\sqrt{3}} \sqrt{F} \right) \frac{U^3} {\lc}
 .
 \label{eq:mhd-eps-v}
\end{equation}
The bracketed factor might be called 
  $ \Ceps^{\text{MHD},v} $
and can be identified with $\Ceps$  in our Eq.~\eqref{eq:eps-phenom}. 
Observationally, 
for the solar wind, $r_A \approx 1/2$ yielding $ F \approx 3 $
  \citep[e.g.,][]{PerriBalogh10-sigc}.
Results from MHD simulations 
   \citep{LinkmannEA17-Ceps, BandyopadhyayEA18-Ceps}\footnote{In both these works $A$ is denoted as $C_{\epsilon,\infty}$ and here we employ double their numerical value for $A$ because of a definitional difference between their $L_\pm$ and our $\ell_f^\pm $.}
indicate that 
    \change[]{$A \approx 0.22 $--$0.27$}{$A \approx 0.5$} 
for situations with zero or moderate mean magnetic field and low or moderate cross helicity, as is relevant to the solar wind.
Using these values we obtain 
  \change[]{$ \Ceps^{\text{MHD},v} \approx 0.05 $ to 0.06}
  {$ \Ceps^{\text{MHD},v} \approx 0.11 $, which is about twice the hydrodynamics estimate of $ \Ceps^\text{hydro} \approx 0.064 $ (via $ A \approx 0.5) $;
   see  \protect\cite{Sreenivasan98} and \protect\cite{PearsonEA04}.
  This gives us values for the prefactor of} Eq.~\eqref{eq:re_taylor} 
  \add[]{of $\frac{\gamma}{C_\epsilon}=\frac{3}{0.11}\approx27$, and of} Eq.~\eqref{eq:re_kolmogorov} 
  \add[]{of $C_\epsilon^{1/3}\approx2$}.  

  The results obtained in this appendix are most relevant for systems governed by the incompressible collisional MHD equations. Thus, application of these results to the nearly collisionless solar wind needs to be undertaken with caution.

    \section{Data Product}
    \label{app:dataset}

Averages of each variable in our dataset are given in Table~\ref{table:dataset_vars}. The dataset (in CSV form), along with metadata describing the variables and the code used to extract and process the data, are available on GitHub\footnote{\texttt{reynolds\_scales\_project} codebase: \url{https://github.com/daniel-wrench/reynolds_scales_project}.} under a 2-Clause BSD License and are archived in Zenodo \citep{wrench_2023_software}. The code has been designed so as to make it relatively simple to apply to data from other missions available in CDAWeb. That is, it should be straightforward to adapt for projects interested in calculating these variables for different heliophysics and space weather environments.

\begin{table}
\centering
\begin{tabular}{||c l r l||}
 \hline
 Symbol & Name & Mean value & Unit \\ [0.5ex] 
 \hline\hline
$SN$ & Sunspot number & 56.3 & - \\ \hline
$M_A$ & Alfv\'en Mach number & 7.36 & - \\ \hline
$M_s$ & Sonic Mach number & 15.31 & - \\ \hline
$\beta_e$ & Electron plasma beta & 0.82 & - \\ \hline
$\beta_p$ & Proton plasma beta & 0.53 & - \\ \hline
$\sigma_c$ & Cross helicity & 0.01 & - \\ \hline
$\sigma_R$ & Residual energy & -0.44 & - \\ \hline
$ R_A$ & Alfv\'en ratio & 0.46 & - \\ \hline
$\cos(A)$ & Alignment cosine & 0.01 & - \\ \hline
$q_i$ & Inertial range slope & -1.68 & - \\ \hline
$q_k$ & Kinetic range slope & -2.64 & - \\ \hline
$Re_{\ts}$ & Reynolds number ($\ts$) & 3,406,000 & - \\ \hline
$Re_{\di}$ & Reynolds number ($\di$) & 330,000 & - \\ \hline
$Re_{\tb}$ & Reynolds number ($\tb$) & 116,000 & - \\ \hline
$f_b$ & Spectral break frequency & 0.25 & Hz \\ \hline
$t_b$ & Spectral break time scale & 14.3 & s \\ \hline
$B_\text{0}$ & Magnetic field magnitude (rms) & 5.49 & nT \\ \hline
$\delta b$ & Magnetic field fluctuations (rms) & 3.83 & nT \\ \hline
$\delta b/B_0$ & Normalized magnetic field fluctuations & 0.71 & nT \\ \hline
$n_e$ & Electron density & 4.18 & cm$^{-3}$ \\ \hline
$n_\alpha$ & Alpha density & 0.14 & cm$^{-3}$ \\ \hline
$T_e$ & Electron temperature & 12.9 & eV \\ \hline
$T_p$ & Proton temperature & 11.0 & eV \\ \hline
$T_\alpha$ & Alpha temperature & 63.8 & eV \\ \hline
$\rho_e$ & Electron gyroradius & 1.78 & km \\ \hline
$\rho_p$ & Proton gyroradius & 63.9 & km \\ \hline
$d_e$ & Electron inertial length & 3.12 & km \\ \hline
$\di$ & Proton inertial length & 134 & km \\ \hline
$l_d$ & Debye length & 0.02 & km \\ \hline
$\lc^\text{fit}$ & Correlation length scale (exp. fit) & 899,000 & km \\ \hline
$\lc^\text{exp}$ & Correlation length scale (1/e) & 942,000 & km \\ \hline
$\lc^\text{int}$ & Correlation length scale (integral) & 880,000 & km \\ \hline
$\ts^\text{ext}$ & Taylor length scale (raw) & 4,770 & km \\ \hline
$\ts$ & Taylor length scale (corrected) & 3,220 & km \\ \hline
$V_0$ & Velocity magnitude (rms) & 439 & km/s \\ \hline
$V_r$ & Radial velocity & 438 & km/s \\ \hline
$\delta v$ & Velocity fluctuations (rms) & 26.2 & km/s \\ \hline
$v_A$ & Alfv\'en speed & 65.5 & km/s \\ \hline
$v_{T_e}$ & Electron thermal velocity & 1490 & km/s \\ \hline
$v_{T_p}$ & Proton thermal velocity & 30.5 & km/s \\ \hline
$\delta b_A$ & Magnetic field fluctuations (Alfven units, rms) & 42.4 & km/s \\ \hline
$z^+$ & Positive Elsasser variable (rms) & 48.9 & km/s \\ \hline
$z^-$ & Negative Elsasser variable (rms) & 48.4 & km/s \\ \hline
\hline
\end{tabular}
\caption{List of the key variables in our Wind data product, comprised of statistics for every 12-hours from 1995-2022. The mean values are for the cleaned 18-year dataset at L1 used in this study. While not shown here, we also provide a few additional variables such the time-scale versions of the length scales, the uncertainty of the Taylor scale, and the amount of missing data for each raw interval. The complete metadata, including the equations used to derive secondary variables such as gyroradii and cross-helicity, can be found in the GitHub README.}
\label{table:dataset_vars}
\end{table}


\bibliography{WrenchReynolds}{}
\bibliographystyle{aasjournal}



\end{document}